\documentclass[12pt,epsfig]{iopart}
\usepackage{epsfig}
\begin{document}

\title[Theoretical Overview]{Theoretical Overview  
QM '04}

\author{Urs Achim Wiedemann}

\address{Department of Physics, CERN, Theory Division,
CH-1211 Geneva 23}

\begin{abstract}
The much wider transverse-momentum range accessible in heavy-ion
collisions at RHIC and at the LHC allows us to disentangle the dynamics
of partonic equilibration from the dynamics of delayed hadronization.
This provides a novel tool for testing the equilibration
mechanisms underlying QCD thermodynamics. Here, I argue, on the
basis of simple formation-time arguments, why this is so, and
I review recent theoretical developments in this context.
\end{abstract}



\maketitle

\section{Hadronization vs. thermalization}
How can one determine, on the basis of the measured hadronic final 
state, whether equilibration processes occur in a medium of 
rapidly decreasing density, and if so, whether they occur in terms 
of partonic or hadronic degrees of freedom? The wide
transverse-momentum range accessible in nucleus--nucleus collisions
at collider energies provides a novel access to this long-standing 
question. 

To see this, consider a parton of high transverse energy
$E_T$, produced in some hard collision: 
if the parton escapes into the vacuum, then it will reduce its 
initial virtuality $Q$ by perturbative parton splitting. After
some time $\sim 1/Q_{\rm hadr}$, its virtuality is degraded
to a hadronic scale $ Q_{\rm hadr} \sim 1$ GeV. Hadronization is
the non-perturbative dynamics of 
the further fragmentation of this multiparton object. 
Numerical estimates for the time scale of hadronization vary 
significantly~\cite{Wang:2003aw,dkmt,Kopeliovich:1990sh}, 
but owing to the Lorentz boost 
to the laboratory frame, they are proportional to the energy, 
$L_{\rm hadr} \sim O(1) \frac{1}{Q_{\rm hadr}} \frac{E_T}{Q_{\rm hadr}}$.

What happens if the hard parton escapes into an infinitely extended 
quark gluon--plasma instead? Because of medium-induced gluon radiation, the 
initial perturbative parton splitting is even more efficient. However, 
the parton cannot hadronize in the dense medium. Instead, after some 
time, its partonic fragments can no longer be distinguished from the 
heat bath: the hard parton is thermalized. To estimate the time scale
$L_{\rm therm}$ for this process, require that the hard parton has 
lost all its energy through medium-induced gluon radiation. According 
to the BDMPS energy loss formula~\cite{Baier:1996sk}, 
$E_T \sim \Delta E = \frac{\alpha_s C_F}{4}
\hat{q} L_{\rm therm}^2$. The partonic thermalization length is
$L_{\rm therm} \sim \sqrt{E_T}$.
%
\begin{figure}[b]\epsfxsize=14.7cm
\centerline{\epsfbox{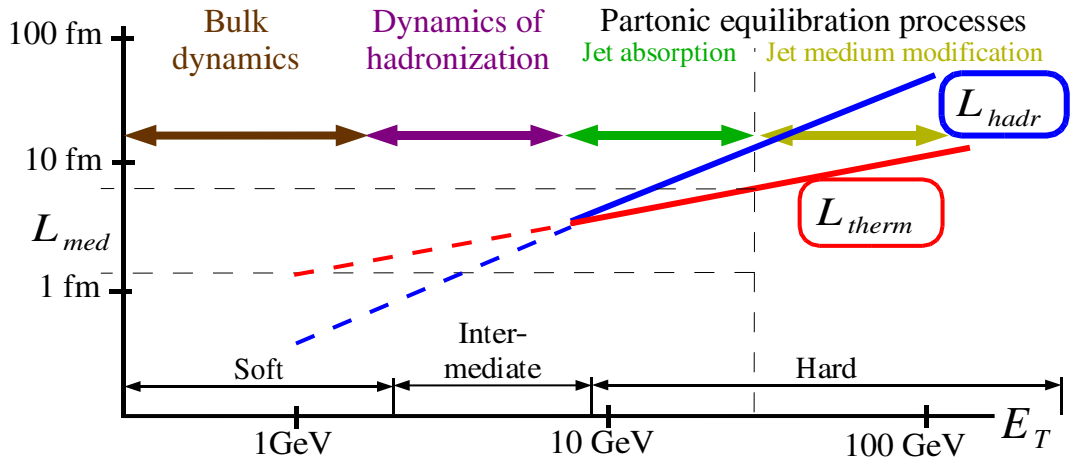}}
\vspace{-4.5cm}
\caption{Comparison of the hadronization time scale
$L_{\rm hadr} = \frac{c}{Q_{\rm hadr}} \frac{E_T}{Q_{\rm hadr}}$,
with the partonic thermalization time scale $L_{\rm therm} =
\sqrt{4/\alpha_s C_F} \sqrt{E_T/\hat{q}}$. For hadronization, 
$Q_{\rm hadr} = 1$ GeV and $c = 2$ to account for multiple
parton branching. For thermalization, 
$\alpha_s\, C_F = 1$ and $\hat{q} = 1$ GeV$^2$/fm. See text
for motivation of parameter values from data.
}\label{fig1}
\end{figure}

The above time estimates are simplified and may be 
improved~\cite{Baier:2000sb}.
They illustrate, however, that for large transverse energies $E_T$, 
perturbative equilibration mechanisms can remain undisturbed
by hadronization over a significant time scale, see Fig.~\ref{fig1}.
Depending on its in-medium pathlength $L_{\rm med}$, the hard parton 
will either be absorbed ($L_{\rm therm} < L_{\rm med} < L_{\rm hadr}$), 
or it has a sufficiently large transverse energy to suffer only 
the onset of equilibration processes ($L_{\rm med} < L_{\rm therm}
< L_{\rm hadr}$). In the latter case, the parton appears as a 
medium-modified jet. For lower transverse energies, there is not
only a competition between the hadronization and the thermalization 
mechanism ($L_{\rm hadr} \sim L_{\rm therm}$). There is also
the possibility that the medium interferes with the dynamics of the
hadronization process ($L_{\rm hadr} \sim L_{\rm med}$).  
For even lower transverse momentum, the hadronization time
scale is determined by the density evolution of the medium, 
which is not accounted for in the above estimates. Only in this 
kinematic ``bulk'' regime may formed hadrons stay in contact with 
the equilibrating medium for a significant duration. I now discuss
in more detail the recent theoretical developments in these 
different kinematic regimes.

\section{Hard $p_T$}
\begin{figure}[h]\epsfxsize=11.7cm
\centerline{\epsfbox{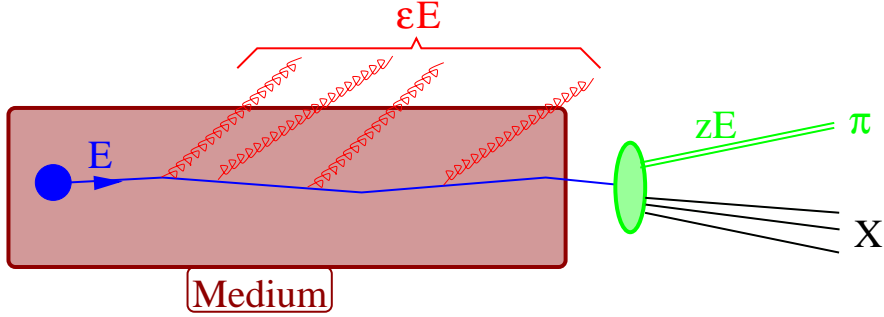}}
\caption{Schematic view of medium-modified leading hadroproduction: 
the hadronization process is assumed to be 
separated in space and time from the medium-modification of
the partonic evolution. This picture implies a trivial particle
species dependence of the nuclear modification factor and applies
at RHIC above $p_T^{\rm parton} \sim 10$ GeV only.   
}\label{fig2}
\end{figure}

For sufficiently hard partons, the hadronization time scale exceeds
the partonic thermalization time scale, $L_{\rm hadr} > L_{\rm therm}$. 
The current understanding of suppressed leading hadroproduction in 
nucleus--nucleus collisions is based on this inequality. One starts from
the assumption that the hard parent partons are produced perturbatively, 
i.e. proportional to the number of binary collisions. The production 
rate of their daughter hadrons, however, is reduced, because some parent 
partons are absorbed in the medium ($L_{\rm med} > L_{\rm therm}$) 
whereas the others ($L_{\rm med} < L_{\rm therm}$) lose an additional 
fraction of their initial energy through medium effects and thus materialise 
in hadrons of lower transverse momentum. As a consequence of this picture 
(see Fig.~\ref{fig2}), all observed high-$p_T$ hadrons have hadronized
{\it outside} the medium. Since hadronization of a parton depends only 
on its identity but not on its history, Fig.~\ref{fig2} implies that
the ratio of different identified leading hadrons stays unchanged with
respect to $p$--$p$ collisions while their absolute yield is suppressed
with respect to the binary scaling assumption. In Au+Au collisions at RHIC,
this is observed in the particle species dependence of the nuclear 
modification factor~\cite{Adler:2003kg,Adams:2003am}, which ceases
for hadronic transverse momenta $p_T^{\rm hadr} > 6$--$7$ GeV 
~\cite{Adams:2003am} (this corresponds to higher partonic momenta 
$p_T^{\rm parton} > 9$--$10$ GeV). This indicates that for the 
observed hadrons, 
$L_{\rm hadr} \sim L_{\rm therm} < L_{\rm med}$ at  $p_T^{\rm parton}
\sim 10$ GeV. This scale compares well with the estimate in 
Fig.~\ref{fig1}. 

%
To account for the partonic medium-modification prior to 
hadronization, we must understand how the parton showers 
associated to hard partonic production processes interact with
a dense medium. This is a difficult problem, which is not solved
completely. All existing 
approaches~\cite{Baier:1996sk,Zakharov:1997uu,Wiedemann:2000za,Gyulassy:2000er,Wang:2001if} indicate
that the leading parton loses additional energy owing to 
medium-induced gluon radiation and that the parton shower broadens 
in transverse momentum. Remarkably, only one property of the medium 
enters these calculations, namely the amount $\hat{q}$ of transverse 
momentum squared that the medium transfers to the parton per unit 
pathlength. Momentum broadening is characterised by 
$\langle k_\perp^2 \rangle \sim \hat{q} L_{\rm med}$~\cite{Baier:1996sk}, 
the additional radiated energy is determined by the characteristic 
energy scale $\omega_c = \frac{1}{2} \hat{q} 
L_{\rm med}^2$~\cite{Baier:1996sk},  and medium effects 
regulate the additional gluon radiation in the infrared on 
a scale $\omega \sim \omega_c/\left( \hat{q} 
L_{\rm med}^3\right)^{2/3}$ ~\cite{Salgado:2003gb}.
Also, the expansion of the medium can be controlled in these
calculations~\cite{Baier:1998yf,Gyulassy:2000gk,Salgado:2002cd}. 
Since the transport coefficient is proportional 
to the density of scattering centres, it decreases in time, 
$\hat{q}(\tau)$. 
The resulting medium-induced radiation spectrum corresponds to
the spectrum of an equivalent static medium in which all kinematic
variables are rescaled by the linear line-averaged transport coefficient
$\bar{\hat{q}} = \frac{2}{L_{\rm med}^2} \int_0^{L_{\rm med}}
\tau\, \hat{q}(\tau)\, d\tau$~\cite{Salgado:2002cd}. In practice, 
this dynamical scaling
law allows a simplified data analysis in terms of static medium
properties; a posteriori, one then translates the extracted static 
average transport coefficient into the realistic dynamical one. 

So far, comparisons of this formalism to data are limited to
the leading hadron spectra and leading back-to-back
correlations. In this case, the reduced energy of the 
leading hadron directly traces the reduced energy of the leading 
parton (see Fig.~\ref{fig2}); the reduction is expected to depend mainly
on the parameter combination $\hat{q}\, L_{\rm med}^2$. The most 
advanced comparisons with data~\cite{Wang:2003aw,Wang:2003mm} reproduce the 
absolute scale of the nuclear modification factor and its centrality 
dependence. This allows us to disentangle information on geometry 
($L_{\rm med}$) and 
density ($\hat{q}$). A satisfactory agreement with data is found for an 
initial energy density of $\epsilon\vert_{\tau_0} =\,\, 
15\, \hbox{GeV}/\hbox{fm}^3$ at initial time 
$\tau_0 = 0.2$ fm/c ~\cite{Wang:2003aw}.
In contrast to other determinations, this is an energy 
density with which a hard parton was {\it interacting}.
The value is consistent with the line-averaged transport coefficient
$\bar{\hat{q}} \sim 1\, \hbox{GeV}^2/\hbox{fm}$ used for the estimate
in Fig.~\ref{fig1}. Based on $\epsilon\vert_{\tau_0}$, one can estimate 
the duration $\tau_{\rm QGP}$ over which the density was above the 
critical energy density $\epsilon_c \sim 0.75\,\,  \hbox{GeV}/\hbox{fm}^3$
predicted by lattice QCD. One finds $\tau_{\rm QGP} = 4 \pm 2 \pm ?$ fm/c,
where the central value is for a one-dimensional Bjorken expansion 
($\tau_{\rm QGP} = \tau_0 \epsilon_{\tau_0}/\epsilon_c$) and $\pm 2$
fm/c is associated to model-intrinsic uncertainties. These stem
from our limited knowledge of the dynamical expansion and evolving 
geometrical distribution of scattering centres, and from significant
uncertainties related to the high-energy approximations employed by
all current calculations (for a quantitative discussion, see 
Refs.~\cite{Salgado:2003gb,Armesto:2003jh}). 
Other systematic errors (e.g. those associated to the perturbative 
calculation of medium-induced gluon radiation) are more difficult to
quantify. 

How can we further test the partonic mechanism of medium-induced
gluon radiation assumed to underlie medium-modified high-$p_T$ 
hadroproduction? Current theory predicts i) the dependence of
medium effects on $L_{\rm med}$, ii) their dependence on the parton
identity, and iii) the relation between leading parton
energy loss and transverse-momentum broadening of the parton shower. 
In my view, the perspectives for refined tests are the following:\\
i) {\it Dependence on geometry}: The approximately quadratic dependence 
on the in-medium pathlength $L_{\rm med}$ is a characteristic feature of 
non-abelian partonic energy loss~\cite{Baier:1996sk}. Typical observables 
sensitive to it 
are the azimuthal anisotropy $v_2(p_T)$ at high $p_T$, the centrality 
dependence of leading hadron spectra and their back-to-back correlations.
In practice, however, uncertainties in modelling the geometry and dynamics 
of the collision, as well as current experimental systematic errors make 
it difficult so far to discriminate even between a linear and a quadratic 
dependences on $L_{\rm med}$~\cite{Drees:2003zh}.\\
ii) {\it Dependence on parton identity:} a) Hard gluons are expected to
suffer larger final-state medium effects than hard quarks since their 
coupling is stronger. Sensitive to this effect is, for example, the ratio
of any pair of leading hadrons that receive different contributions 
from parent quarks and parent gluons (e.g. the $p/\bar{p}$ ratio).
In practice, however, an analysis of these measurements is complicated by 
the experimental limitations on high-$p_T$ particle identification, and by
the uncertainties in the vacuum fragmentation functions at large momentum
fraction $z$~\cite{Zhang:2002py}. 
At sufficiently high luminosity and/or centre-of-mass
energy, a more promising test of gluon propagation in matter may be
$\gamma$-jet or three-jet ($q\, q\, g$) events where tagged leptonic 
or hadronic high-$p_T$ information signals the presence of a gluon.
b) Massive quarks are expected to show smaller medium modifications
than massless ones and this should show up in the leading charmed and 
beauty hadrons~\cite{Dokshitzer:2001zm,Armesto:2003jh,Djordjevic:2003zk,Zhang:2003wk}. However, for massive quarks, $L_{\rm hadr}$ is reduced 
since massive quarks are slower, and $L_{\rm therm}$ is increased 
since they lose less energy. As a consequence (see Fig.~\ref{fig1}), 
the transverse energy scale where $L_{\rm hadr} \sim L_{\rm therm}$ and 
partonic energy loss calculations start to apply may lie at the
upper end of the current experimental data, above 
$p_T^{\rm hadr} \sim 10$ GeV. \\
iii) {\it Beyond leading hadron spectra:} Calculations predict how the 
energy lost by the leading parton is redistributed in transverse phase 
space~\cite{Wiedemann:2000tf,Salgado:2003gb}. Ideally, 
this can be tested by reconstructing the total energy 
associated to the initial parent parton, i.e. by measuring the 
``medium-modified jet''. The corresponding jet production cross section 
is expected to follow binary scaling, but the jet shape is broadened 
and the jet multiplicity is softened and increased~\cite{Salgado:2003rv}. 
The main problem 
of calorimetric jet measurements is the underlying high-multiplicity 
background. At present, this is studied intensively for the heavy-ion 
programme at the LHC, and first jet observables are known, which are 
sensitive to transverse-momentum broadening but insensitive to the 
low-$p_T$ background~\cite{Salgado:2003rv}. Where calorimetric 
measurements are not feasible, a pragmatic widely adopted alternative 
is the study of 
jet-like near-side particle correlations associated to high-$p_T$ 
trigger particles~\cite{Pal:2003zf}. Finding in this way the lost 
remnants of jets is 
certainly an important qualitative support of the assumption of 
medium-induced final-state energy loss. However, it is difficult
to compare the trigger selection of a special subclass of all jets 
(namely those with a particularly energetic leading hadron) with
current parton energy 
loss calculations without significant further assumptions. This is 
so because theory is more reliable for energy distributions
than for multiplicity distributions (which require a more detailed
knowledge of hadronization), and because the same leading-particle
trigger selects significantly different jet classes with different
average $E_T$ depending on the a priori unknown medium modification.
%
\begin{figure}[h]\epsfxsize=14.7cm
\centerline{\epsfbox{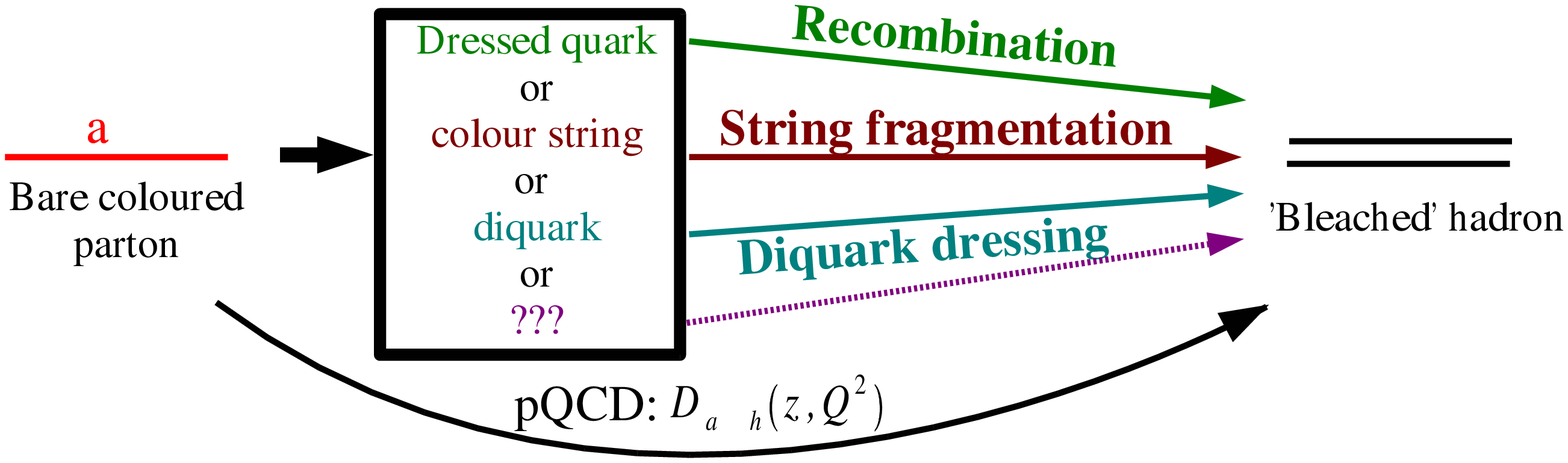}}
\vspace{-6.5cm}
\caption{Schematic view of the dynamics of hadronization. In
nucleus--nucleus collisions, the medium allows a test of the
non-perturbative long-range part of parton fragmentation if 
the transverse momentum is not too large. 
}\label{fig3}
\end{figure}
%
\section{Intermediate $p_T$}

Hadronization is the non-perturbative process that bleaches coloured 
partons into sets of hadronic states. As of today, its dynamics 
remains a black box (see Fig.~\ref{fig3}) which is filled differently 
by different hadronization models. A satisfactory theoretical 
understanding exists only at sufficiently high $p_T$ when factorization 
theorems are applicable. This black box can then be bypassed by a 
simple mapping, the fragmentation function $D_{h/q}(z,Q^2)$. In this 
case, the only dynamical information accessible about hadronization 
is the dependence on the scale $Q^2$ at which perturbative physics 
is interfaced with $D_{h/q}(z,Q^2)$. This limits the possibility to study 
{\it how} hadronization proceeds. 

The same mapping $D_{h/q}(z,Q^2)$ is applied in the current description 
of medium-modified hard processes (see Fig.~\ref{fig2}). As a consequence, 
the particle ratio of leading protons/pions is predicted to trace the 
ratio of the corresponding fragmentation functions, $p/\pi \sim
D_{p/q}(z,Q^2)/D_{\pi/q}(z,Q^2) \sim 0.1$. However, in Au+Au collisions
for $p_T^{\rm hadron} < 7$ GeV, one observes a marked 
deviation~\cite{Adler:2003kg,Adams:2003am}
from this perturbative ratio, which increases with centrality
and reaches unity. Thus, perturbative fragmentation breaks down in Au+Au
collisions although it is known to provide a satisfactory description
of the particle species dependence of elementary $e^+ e^-$ or hadronic
collisions in the same intermediate-$p_T$ regime. The scale 
$p_T^{\rm hadr} < 7$ GeV where this happens lies in the range where 
$L_{\rm med} \sim L_{\rm hadr} \sim L_{\rm therm}$ (see Fig.~\ref{fig1}).
This suggests that the much larger spatial extension of dense matter 
in nucleus--nucleus collisions interferes with the hadronization
process at higher $p_T$. The medium becomes a tool for testing the
dynamics of hadronization. From this perspective, the theoretical 
challenge at intermediate $p_T$ is to discriminate, e.g. between the different 
proposed prehadronic objects through which hadronization may occur. 

At present, such a program has at best started. Different models, 
which focus on hadronic spectra at intermediate $p_T$ 
i) interpolate between soft hydrodynamic evolution and hard high-$p_T$ 
processes~\cite{Hirano:2003pw}, including known nuclear 
effects~\cite{Gyulassy:2000gk}, ii) model the dynamics of string 
fragmentation~\cite{Cassing:2003sb}, or iii) aim to account for data
in terms of the recombination of dressed 
quarks~\cite{Voloshin:2002wa,Hwa:2002zu,Fries:2003vb,Greco:2003xt}. 
In all these approaches, 
the particular non-perturbative model components become numerically
unimportant outside the intermediate-$p_T$ regime above $p_T^{\rm hadr} 
\sim 7$ GeV~\cite{Fries:2003vb}. In particular, recombination models
offer an explanation for the anomalous particle species dependence
at intermediate $p_T$ in terms of a simple prehadronic picture. For
example, they emphasise that the elliptic flow of different baryons 
and mesons falls on an universal curve $\frac{1}{n} v_2(p_T/n)$ if 
rescaled by the number $n$ of valence quarks~\cite{Voloshin:2002wa}. 
This may be interpreted as a
remnant of a common underlying partonic flow, which recombines differently 
for different hadron spectra~\cite{Molnar:2003ff}. It opens the exciting 
possibility that the
partonic flow saturates at a rather small value of $\sim 7 \%$ but 
that it is amplified by a factor 2--3 in the hadronization process. 
Whether this suggestive counting rule can result from a dynamical 
context (which should explain, for instance, the fate of gluonic degrees of 
freedom) and how it can be discriminated, e.g. from models of string
fragmentation, remains to be established. 

In short, the strongest argument for the novel physics potential at 
intermediate $p_T$ is the coincidence of scales $L_{\rm med} \sim 
L_{\rm therm} \sim L_{\rm hadr}$, which indicates the interference
of the hadronization process with the medium. The data show clear 
signals of such an interference~\cite{Adler:2003kg,Adams:2003am}.

\section{Soft $p_T$}

Hadronic spectra at high and intermediate $p_T$ differ significantly
from thermal and chemical equilibrium distributions, but theory 
relates their medium-dependence dynamically to the onset of partonic 
equilibration and medium-dependent hadronization. 
At soft $p_T$ ($p_T < 2$ GeV), the challenge is the
inverse: many bulk observables show (approximate) equilibrium
distributions and the task is to establish to what extent this is
the consequence of equilibration processes.
   
In particular, particle ratios do not show significant deviations 
from chemical equilibrium and can be described by the model of statistical 
hadronization in terms of a temperature ($T_{\rm ch} \sim 170$ MeV)
and a baryochemical potential ($\mu_{\rm B} \sim 40$ MeV at 
$\sqrt{s_{NN}} = 200$ GeV)~\cite{Braun-Munzinger:2003zd}. 
However, the observed equilibrium does not automatically imply a dynamical 
equilibration mechanism, since statistical particle production according to 
the principle of maximum entropy is sufficient to motivate the underlying
model. A dynamical origin of the observed equilibrium distribution 
still remains an intriguing possibility~\cite{Braun-Munzinger:2003zz}, 
in particular because the extracted temperature and baryochemical 
potential lie close to the phase-transition line of equilibrated 
lattice QCD~\cite{Fodor:2001pe,deForcrand:2002ci,Allton:2002zi,D'Elia:2002gd}. 
But to establish a
firmer link between the phase-transition line to a QGP and the 
statistical description of hadron yields is most likely to require
a better understanding of the hadronization process. 

Moreover, combinations of the model of statistical hadronization
with hydrodynamically motivated parametrisations of the bulk 
dynamics at kinetic freeze-out (the so-called ``blastwave models'') 
account well for the particle-species dependence of soft $p_T$ 
hadronic spectra and their azimuthal dependence 
$v_2(p_T)$~\cite{Retiere:2003kf}. [Intriguingly, this implies that 
at soft $p_T$ the species dependence of elliptic flow 
is a mass effect, while it is argued to be a 
baryon/meson effect at intermediate $p_T$~\cite{Voloshin:2002wa}.] 
These combinations also account for 
the space-time structure reflected in identical particle correlations,
and in particular for the recently measured azimuthal dependence of 
these correlations, which trace the space-time picture of elliptic flow.
The model parameters extracted from fits to data suggest that the
kinetic freeze-out temperature lies significantly below the chemical
one ($T_{\rm ch} - T_{\rm kin} \sim 70$ MeV $\sim O(T_{\rm kin})$)
and that the system shows a strong collective transverse expansion
$v_T > \frac{1}{2} c$. They also suggest a rather short
freeze-out time of $\tau_f \leq 10$ fm/c. [This
bound is based on the assumption of Bjorken longitudinal expansion.
If the longitudinal expansion is slower as, for instance, suggested by 
the description of $dN/dy$ in terms of Landau hydrodynamics~\cite{murray}, 
the freeze-out time can then be significantly 
larger, see Fig. 1 of~\cite{Wiedemann:1999ev}.] 
However, blastwave models are only (physically motivated) 
parametrisations of data. They   
do not address the question of which microscopic dynamics leads to the 
observed distributions.

Hydrodynamics, based on the assumption of perfect local thermal and 
chemical equilibrium, is a dynamical explanation. It can account for
hadronic transverse~\cite{Hirano:2003yp,Kolb:2002ve} and 
longitudinal~\cite{Hirano:2003yp} single-particle spectra and their azimuthal 
dependence~\cite{Kolb:2001qz}. In contrast, hydrodynamic calculations do not 
reproduce the identical two-particle correlations~\cite{Heinz:2002un}. 
It remains unclear whether this is a problem of hydrodynamics 
itself~\cite{Heinz:2002un}, or of the freeze-out condition with which 
hydrodynamics is 
interfaced~\cite{Sinyukov:2002if,Akkelin:2001wv,Tomasik:2002hs}. 
The reason is that HBT measurements test the surface of 
last scattering, which can change significantly depending on the 
freeze-out condition while the large mean-free path makes hydrodynamical 
assumptions questionable. On the other hand, a clearly model-intrinsic 
problem of hydrodynamics is to understand microscopically {\it how} 
a hydrodynamic behaviour comes about~\cite{Molnar:2001ux}. Simulations indicate
that for the parton densities and partonic cross sections expected at 
RHIC, the pressure built up in a peripheral collision can result in
an elliptic flow of $\sim 5 \%$ at most. This falls short of the 
$\sim 15\%$ observed on the hadronic level. Here, the novel provocative 
suggestion from recombination models is that this small partonic $v_2$ 
may be almost sufficient since it gets amplified in the hadronization 
process~\cite{Molnar:2003ff}. It is important to establish whether 
this is a viable 
explanation, since this would also strongly affect the understanding 
of $v_2$ at $p_T < 2$ GeV. If true, it implies that insight into the 
hadronization process is indispensable for understanding whether and 
how hydrodynamics works on the partonic level.  

\section{Saturation physics - A complementary view}

The discussion of Fig.~\ref{fig1} is not the only way to attach
fundamental physics questions to the transverse momentum scale.
Saturation physics provides an alternative view. Its starting point 
is the observed growth of parton distribution functions at small $x$. 
On general grounds~\cite{McLerran:2003yx}, this growth must be tamed 
eventually by density 
effects and it will saturate for densities of order $\sim 1/\alpha_s$. 
In this regime, perturbation theory in a high-density background 
allows for the derivation of density-modifications to the 
small-$x$ BFKL 
evolution~\cite{Balitsky:1995ub,Kovchegov:1999yj,Jalilian-Marian:1997dw}. 
As a consequence, the unintegrated gluon distribution is depleted 
(saturated) up to a saturation scale $Q_s$
~\cite{McLerran:1993ka,Jalilian-Marian:1996xn,Golec-Biernat:1999qd}. 
This scale can grow perturbatively
large (possibly $Q_s^2 \sim (1$--$2 {\rm GeV})^2$ at RHIC), and it
increases with the nuclear extension ($Q_s^2 \sim A^{1/3}$) and
for small $x$ or increasing rapidity $y$ 
($Q_s(y) \sim \Lambda e^{\alpha_s\, c\, y}$, $c \sim O(1)$).
In addition to the perturbative high-$p_T$ region in
which density effects are negligible, saturation physics identifies
two transverse-momentum regimes with qualitatively different physics:

{\it Soft $p_T$: saturation region.} 
Since the initial parton density is depleted, one may expect that less 
partons will be produced in the collision. It has been argued that
this can explain the low event multiplicities at RHIC~\cite{Kharzeev:2000ph} 
and that it provides a fair description of bulk observables such as rapidity 
distributions and their centrality dependence~\cite{Kharzeev:2001gp}. 
However, a significant
model-dependent input is required to interface saturation physics 
with soft bulk observables, and models based on other equally
fundamental physics assumptions are equally successful, 
see e.g.~\cite{Hirano:2003yp}.

{\it Intermediate $p_T$: scaling window.}
The low-$k_T$ ($k_T^2 < Q_s^2$) gluons of a depleted gluon distribution  
are the dominant source of high-$k_T$ gluon production of  during
the small-$x$ evolution. Hence, small-$x$ evolution  
reduces the growth of the gluon distribution above the saturation 
scale, in the so-called scaling window $Q_s^2(x_{\rm evolved})
< k_T^2 < Q_s^4(x_{\rm evolved})/Q_s^2(x_{\rm initial})$
~\cite{Iancu:2002tr,Mueller:2002zm,Albacete:2003iq}. Since
the saturation scale grows with $x$, this scaling window opens as 
a function of increasing cms energy or rapidity. One therefore
expects in both hadron--nucleus and nucleus--nucleus collisions 
a suppressed hadroproduction at intermediate $p_T$ iff 
the saturated gluon distribution is evolved sufficiently 
far in $x$~\cite{Kharzeev:2002pc}. At mid-rapidity at RHIC, this 
scenario is ruled out by the observed opposite centrality dependence of 
the nuclear modification factor in d--Au and Au--Au
~\cite{Arsene:2003yk,Adler:2003ii,Back:2003ns,Adams:2003im}. 
This does not exclude the parton 
saturation as a viable explanation in the soft regime $p_T^2 < Q_s^2$ 
at RHIC mid rapidity. It just shows that the saturated gluon distribution 
was not yet evolved sufficiently far in $x$ for the scaling window 
to open.

The observed disappearance of the Cronin enhancement in d--Au at 
forward rapidity~\cite{debbe} was predicted as a consequence of the 
small-$x$ evolution of a saturated gluon 
distribution~\cite{Albacete:2003iq,Kharzeev:2003wz}. 
In the context of saturation physics, this signals
that the scaling window has opened because of the small-$x$ evolution.
However, the agreement is at best qualitative: the 
studies~\cite{Baier:2003hr,Albacete:2003iq,Kharzeev:2003wz}  
calculate, on the basis of a kinematically simplified partonic cross
section, a ratio of gluon spectra; neither the validity of this
simplification, nor the effect of neglecting quarks has been discussed 
so far. Also, alternative explanations are not fully 
explored yet. In general, if multiple scattering leads to an effective
redistribution of hadronic yields not only in transverse but also in
longitudinal momentum, this could then deplete the ratio $R_{\rm dAu}$ 
at forward rapidity equally well. Calculations of $R_{\rm dAu}$ based on
multiple scattering~\cite{Vitev:2003xu,Accardi:2003jh} did not confirm
this effect, but predicted a Cronin enhancement at forward rapidity
instead. However, these calculations are based on a high-energy approximation 
in which a transverse-momentum transfer does not degrade longitudinal 
momentum and where only elastic multiple-scattering contributions 
are considered. Thus, it is still conceivable that refinements change 
the result of conventional Cronin calculations significantly.

 If the interpretation of the rapidity dependence of $R_{\rm dAu}$
in terms of saturated gluon distributions stands further tests,  
then this has far-reaching consequences. At RHIC, one expects
further observable traces, in particular for open charm 
production~\cite{Eskola:2003fk,Kharzeev:2003sk} and possibly for 
dileptons~\cite{Jalilian-Marian:2004er}. For the LHC, a confirmed onset 
of parton saturation at forward RHIC rapidity implies that the 
medium attained in Pb--Pb collisions at the LHC differs qualitatively 
from that at RHIC, since it corresponds to largely evolved saturated 
parton densities. 
In the long term, this finding would give further support to
the already strong case for a dedicated $e$--$A$ experimental program.

\section{Conclusion}

In summary, the heavy-ion programs at RHIC and the LHC are in the
fortunate situation that hadronization and partonic thermalization
length scales become comparable in the very range of
in-medium pathlength and transverse momentum accessible at collider 
energies (see Fig.~\ref{fig1}). This suggests that the much wider 
$p_T$-range at RHIC and LHC provides a novel tool to disentangle 
the two recurrent dynamical features of heavy-ion collisions: the 
dynamics of partonic thermalization and the dynamics of medium-dependent 
hadronization. I argued why the main open questions at soft $p_T$ 
are linked to our poor understanding of hadronization and freeze-out, 
and how the study of data at intermediate $p_T$ may improve this 
understanding. At high $p_T$, I discussed the possibility to 
study {\it how} an out-of-equilibrium parton -- undisturbed by hadronization 
--evolves towards partonic equilibrium.
This is not only an important step towards QCD thermodynamics. 
A refined theoretical understanding of medium-modified hard probes
also allows for more detailed tests of the dense bulk
matter with strongly interacting penetrating probes. 

Clearly, the RHIC-ness of scales accessible at collider energies is
not exhausted by this discussion. Varying other scales such as the 
$A$-dependence and energy dependence of nucleus--nucleus collisions
provides complementary information, for example for a precise scan
of energy density and baryochemical potential. Also, other
views on the wide $p_T$ range, such as saturation physics, may turn
out to be more fruitful in the future. Whatever this future may be,
progress in the strong interactions has always been based on the 
interplay of theory and experiment. The number of accessible scales 
and fundamental concepts illustrates clearly that at present one of 
the most active and most diverse fields for this interplay is the 
study of nucleus--nucleus collisions.

\vspace{0.5cm}


\end{document}